\documentclass{article} % For LaTeX2e
\usepackage{nips13submit_e,times}
\usepackage{hyperref}
\usepackage{url}

\usepackage{mathtools}
\usepackage{subcaption}
\usepackage[justification=centering]{caption}
\usepackage{graphicx}

\title{Exploiting Direct And Indirect Information\\
 For Friend Suggestion In ZingMe}

\author{
Kien Duy Nguyen \\
School of Information and Communication Technology \\
Hanoi University Of Science and Technology \\
Hanoi, Vietnam \\
\texttt{duykienvp@gmail.com} \\
\and
Tuan Pham Minh \\
VNG Corporation \\
Hanoi, Vietnam \\
\texttt{tuanpm@vng.com.vn} \\
\and
Quang Nhat Nguyen \\
School of Information and Communication Technology \\
Hanoi University Of Science and Technology \\
Hanoi, Vietnam \\
\texttt{quang.nguyennhat@hust.edu.vn} \\
\and
Thanh Trung Nguyen \\
VNG Corporation \\
Hanoi, Vietnam \\
\texttt{thanhnt@vng.com.vn} \\
}

\nipsfinalcopy % Uncomment for camera-ready version

\begin{document}

\maketitle

\begin{abstract}
Friend suggestion is a fundamental problem in social networks with the goal of assisting users in creating more relationships, 
and thereby enhances interest of users to the social networks. 
This problem is often considered to be the link prediction problem in the network. 
ZingMe is one of the largest social networks in Vietnam. 
In this paper, we analyze the current approach for the friend suggestion problem in ZingMe, 
showing its limitations and disadvantages. 
We propose a new efficient approach for friend suggestion that uses information from the network structure, 
attributes and interactions of users to create resources for the evaluation of friend connection amongst users. 
Friend connection is evaluated exploiting both direct communication between the users 
and information from other ones in the network. 
The proposed approach has been implemented in a new system version of ZingMe. 
We conducted experiments, exploiting a dataset derived from the users' real use of ZingMe, 
to compare the newly proposed approach to the current approach and some well-known ones for the accuracy of friend suggestion. 
The experimental results show that the newly proposed approach outperforms the current one, 
i.e., by an increase of 7\% to 98\% on average in the friend suggestion accuracy. 
The proposed approach also outperforms other ones for users who have a small number of friends
with improvements from 20\% to 85\% on average.
In this paper, we also discuss a number of open issues and possible improvements for the proposed approach.
\end{abstract}

\section{Introduction}

Online social networks bring people a new way to receive, to exchange and to share information. 
In order to attract and keep interest of users, one important problem in social networks 
is how to make virtual society of users bigger and their relationships closer. 
Relationships can be friendships, profession links, collaborations and so on. 
Suggesting new relationships for users is one way of tackling this. 
We can define this problem as follows: given a snapshot of the social network at time $t$ and for each user, 
we want to predict new relationships that will appear in the future \cite{liben03}. 
This problem is often considered as link prediction or link recommendation problem:  
given a snapshot of a network at time $t$, we want to output a predicted list of edges that will be added to the network in the future. 

There is one question we should consider for friend suggestion problem: can new friendships of users be predicted accurately based on information in the network \cite{back11}? Many reasons may cause new friendships between users. For instance, they can meet each other in a party and then make friends in the network. However, we can guess that when they come to a party, they properly have some mutual friends or work for same company. If we can obtain that information, we can have foundation to believe in success of predictions. 

Another challenge is a massive class skew \cite{rattigan05}. Data from ZingMe shows that only 0.02\% of possible friendships are actually established during the period of data collection. This causes a difficulty in distinguishing established friendships from non-established ones.

ZingMe is one of the largest social networks in Vietnam with about 8 millions active users per month.
In ZingMe, we call this problem friend suggestion problem because relationships among users are friendships.
ZingMe currently has an approach to solve this problem but it has some limitations and disadvantages.
Although current approach uses some important information sources to access ability to create friendship between two users,
there are still other ones ignored.
In addition, it uses only information between two users, wastefully discarding information from other users in the network
which can also affect to progress of creating friendship between them.
This leads to the need for a new approach that can efficiently exploit information sources of this social network.

Our work presents a novel approach for friend suggestion problem on social networks that can effectively exploit information between two users and information from other users in the network in suggestion process. Our approach also leverages many different knowledge sources such as network structure, attributes of vertices and attributes of edges. Our work also brings users of ZingMe a new friend recommender system that captures better their interests in the network. We also anonymize and give our dataset available to research community. 

The rest of this paper is organized as follows. Section 2 presents current approach deployed for ZingMe. Our proposed approach is presented in Section 3. Some experiments and discussions are given in Section 4. Related works are summarized in Section 5. Finally, Section 6 concludes this paper and shows some future directions.

\section{The Friend Suggestion Approach Currently Used in ZingMe}
In ZingMe, each user may have a list of friends, a list of IP addresses he used to log in to ZingMe, a list of schools and a list of companies he have studied or worked.
%These user information is stored in back-end services based on ZDB key-value storage \cite{zdbthanhnt} for high performance and low latency when accessing.
Currently, a approach deployed to solve friend suggestion problem in ZingMe that uses following features between two users: 
\textit{number of mutual friends}, \textit{number of mutual schools}, \textit{number of mutual companies} and \textit{number of mutual IP addresses}. 

For a target user $u$ and a candidate $v$, the current approach builds a graph to assess ability to form friendship from $u$ to $v$. 
Given $N_{u}$ as number of friends of $u$, the graph includes four vertices and edges presenting strength of relationships amongst those vertices.
The vertices have $t_{1}$, $t_{2}$, $t_{3}$, $t_{4}$ as weights 
when $u$ and $v$ have or do not have mutual friends, mutual schools, mutual companies, mutual IP addresses, respectively.
$t_{i} = 1$ or $t_{i} = w_{i} > 1$ when $u$ and $v$ have or do not have mutual information for feature $i$.
A vertex $i$ is "turned on" if $u$ and $v$ have mutual information for that vertex. Especially, vertex $1$ is always turned on.

Edge $(i, j)$ has weight $e_{i, j}$ that indicates weight when both of $t_{i}$ and $t_{j}$ are turned on, and $e_{i, j}$ is turned on, otherwise, is equal to $1.0$. 
For example, if we think that for two users, having mutual companies and mutual IP addresses is more important than having mutual schools and IP addresses,
 we may set $e_{3, 4} > e_{2, 4}$. Given $n$ as number of turned on edges, function for accessing ability to make friends between $u$ and $v$ as follows:
 \begin{equation}
 \textit{score}_{u, v} = n * \sum_{i, j = 1}^{4} e_{i, j} * t_{i} * t_{j}
 \end{equation}
 
 We notice that the current approach only uses some knowledge sources in the network for friend suggestion 
 and neglects many potential sources such as groups of users or interactions of users. 
 It also creates strong ties among features through edges of the graph, thereby creating difficulties in adding more features.
 In addition, it only exploits direct information between two users, discarding information from other ones in the network.

 \section{The Proposed Friend Suggestion Approach}
 
 \subsection{Formal Definitions}
Social network is modelled as directed graph $G = (V, E)$ in which each vertex $u \in V$ represents a user in the network 
and each edge $e_{u, v} \in E$ represents relationship: $v$ is a friend of $u$. 
Because of symmetry in friendship in social network ZingMe, if $e_{u, v}$ exists, $e_{v, u}$ also exists. 
Weight $c_{u, v}$ represents strength of friendship between $u$ and $v$. 
$t_{u, v}$ or $t_{e}$ is the time when edge $e_{u, v}$ is established. 

For times $t, t_{1}, t_{2}$, let $G[t] \subseteq G$ and $G[t_{1}, t_{2}] \subseteq G$ consist of edges with $t_{e} \leq t$ and $t_{1} \leq t_{e} \leq t_{2}$, respectively. 
The link prediction problem for social networks can be defined as follow: given times $t_{0} < t'_{0} \leq t_{1} < t'_{1}$, based on $G[t_{0}, t'_{0}]$,
 we want to output a list of edges which are not in $G[t_{0}, t'_{0}]$ and are predicted to present in $G[t_{1}, t'_{1}]$ \cite{liben03}.
 
 \subsection{Knowledge Sources}
 The proposed friend suggestion approach uses knowledge sources: 
 \textit{network structure}, 
 \textit{attributes of vertices}, for example, age, sex or education
 and \textit{attributes of edges}, for example, interactions when a user comments on a picture, likes a page or sends a message
 
Specific features are derived from these knowledge sources with assumption that 
if two vertices are not connected, the more mutual information they have, the more probability to be connected they have, 
and if they are already connected, it represents strength of their connection which is greater when they have more mutual information. 
Simultaneously, type and reliability of data are also considered when choosing these features.
 
For the graph structure, \textit{the number of mutual adjacent vertices} is the feature derived and equal to $|N_{u} \cap N_{v}|$. 
Data from ZingMe and Facebook \cite{back11} also show that more than a half of new friendships are established from users having mutual friends. 

For the attributes of vertices, selected features are \textit{number of mutual schools}, \textit{number of mutual groups} and \textit{number of mutual IP addresses}.

For the attributes of edges, the feature chosen is \textit{number of "mutual interactions"}. For instance, if two users like same picture, that can be considered as a mutual interaction between them. There are many types of mutual interactions such as users are tagged in same pictures, comment on same posts, and it may be better to consider each mutual interaction as an individual feature but they are considered as one because of computing capability of real system. 

Our approach reuses some features from current approach, including number of mutual adjacent vertices, number of mutual schools and number of mutual IP addresses.
It also adds features derived from groups and interactions of users, which were neglected in the current approach. 
It eliminates the feature relating to companies of users because of unreliability of this feature. 
In ZingMe, users provide names of their companies by free text without any standardization.
By manually examining content of those names, we found them unreliable.

\subsection{The Proposed Approach}
The proposed approach for predicting new connections for a vertex $u$ is based on accessing ability to form connections between $u$ and other ones. Given a candidate $v$, it used direct information between $u$ and $v$, along with indirect information from other vertices. These types of information are obtained from \textit{affinity} between two vertices.
\begin{figure}[h]
 \begin{center}
 \includegraphics[scale=0.4]{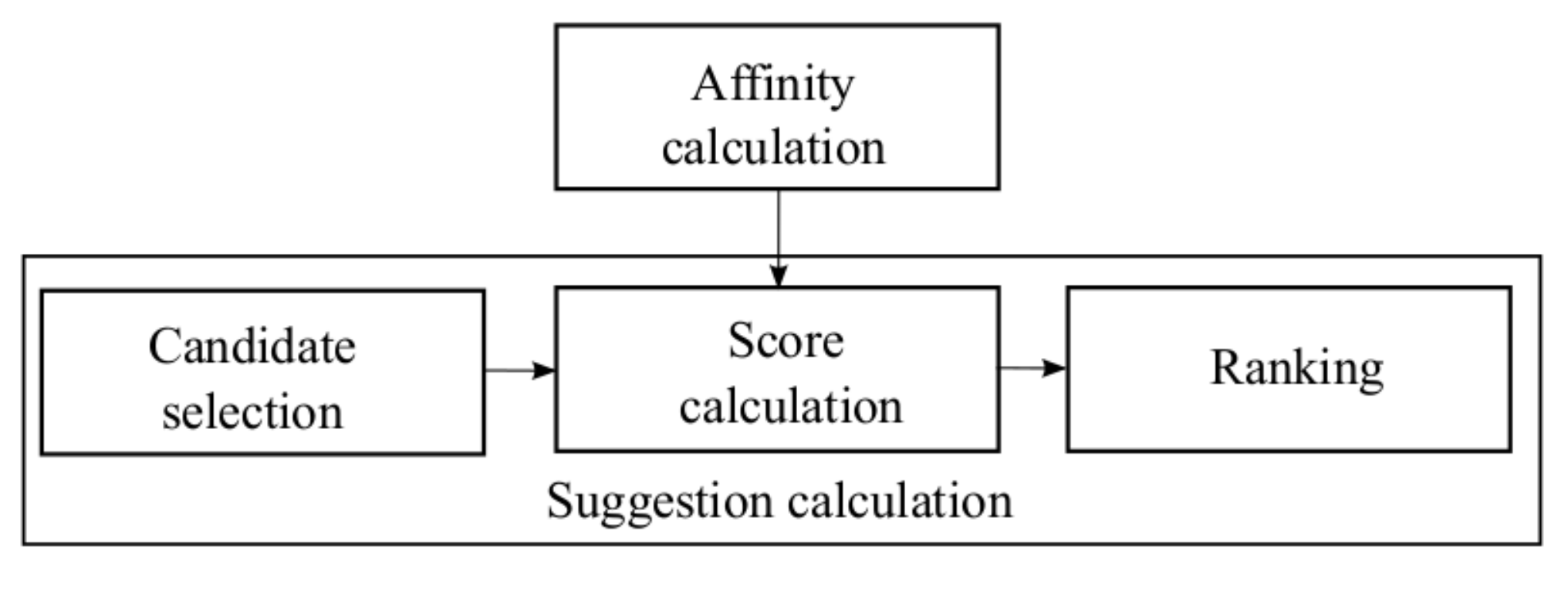}
 \caption{Proposed approach diagram} 
 \label{fg:proposedapproach}
 \end{center}
\end{figure}

As illustrated in Figure~\ref{fg:proposedapproach}, proposed approach consists of two components: \textit{affinity calculation} and \textit{suggestion calculation}. 
Suggestion calculation component includes three phases to calculate suggestions for a target vertex $u$: 
candidate selection phase finds potential candidates to form edges to $u$; 
score indicating ability to form edge to $u$ of each candidate is calculated in score calculation phase; 
ranking phase ranks candidates to output a suggestion list for $u$.  

\subsubsection{Affinity Calculation}
Affinity of $u$ to $v$, $\textit{aff}_{u, v}$ indicates how $u$ is interested in or affined to $v$ and is used as $c_{u, v}$ on the graph. 
It is noted that $\textit{aff}_{u, v}$ may not be equal to $\textit{aff}_{v, u}$. 
In order to calculate $\textit{aff}_{u, v}$, features are used based on the hypothesis that the more mutual information $u$ and $v$ have, the higher $\textit{aff}_{u, v}$ is. In addition, each feature has its own level of influence to $\textit{aff}_{u, v}$, i.e., weight. The computation of $\textit{aff}_{u, v}$ is defined as follows: 
\begin{equation}
\label{eq:aff}
\textit{aff}_{u, v} = \sum_{i = 1}^{\textit{Nf}} w_{i} * log(S_{i} + 1)  \\
\end{equation}
in which $\textit{Nf}$ is the number of features, $w_{i}$ is the weight for feature $i$ and $S_{i}$ is the value of feature $i$. 
The logarithm function is used for normalization because for different features, their values $S_{i}$ may belong to very different ranges. 
For example, the number of mutual schools between two users is often smaller than 3 while the number of mutual interactions may get value at tens or hundreds. 
The logarithm function helps features get in closer ranges. 

\subsubsection{Suggestion Calculation}
\textbf{Candidate Selection} 
The goal of the candidate selection phase is to select potential candidates to connect to target vertex $u$.
This issue arises when number of vertices is very large. Calculating suggestion scores from $u$ to all vertices is infeasible. 
The selection method must satisfy two requirements: selected candidates should have high potentiality to establish connection from $u$ 
and should be suitable for calculation method to ensure acceptable computation time. 

Given $C_{u}$ as candidate set of $u$, $L$ as maximum number of candidates and a threshold $\mu$, 
the candidate selection method sorts all vertices having path of length two to $u$ in descending order by the number of mutual neighbours. 
Then, for each vertex $v$ in this sorted list, it adds $v$ to $C_{u}$ if $|C_{u}| < L$  and $\mu \leq |N_{u} \cap N_{v}|$ . 

In social networks, candidates are friends of friends of $u$ but not friends of $u$. 
This comes from the fact that more than a half of new friendships in ZingMe and Facebook \cite{back11} are friends of friends.
 We sort candidates by number of mutual friends with assumption that users with more mutual friends with $u$ may be easier to create new connections to $u$. 
 The number of mutual friends and the number of candidates is also limited to improve prediction efficiency and to reduce computation time.
 
 \textbf{Score Calculation}
 $\textit{score}_{u, v}$ indicates ability (i.e., the system's confidence level) to establish connection from $u$ to $v$. 
 It is calculated from two knowledge types: direct information between $u$ and $v$, $\textit{aff}_{u, v}$, and indirect information from others on the graph, $\psi_{u, v}$. 
 The computation $\textit{score}_{u, v}$ is defined of  as follows:
\begin{equation}
\label{eq:score}
 \textit{score}_{u, v} = w_{d} * \textit{aff}_{u, v} + w_{i} * \psi_{u, v}
 \end{equation}
 in which $w_{d}$ and $w_{i}$ are weights for direct and indirect information, indicating the level of their influence to the formation of the connection between $u$ and $v$.
Calculation method for $\psi_{u, v}$ should leverage properties of connections of other vertices that can affect ability for $u$ to connect to $v$. These properties for a pair of vertices include number and quality of paths between them such as length of a path or weight of each connection. 
 $\psi_{u, v}$  calculation can be turned into proximity measures in graph. Random walk with restart (RWR) is a well-known approach for this problem  \cite{tong06}. 
 
 RWR can be defined as following equation:
\begin{equation}
r_{i} = (1-\alpha)*r_{i}*A + \alpha * e
 \end{equation}
 in which $A$ is transition matrix, $r_{i}$ is stationary distribution of RWR at iteration $i$, $e$ is unit vector with $e_{u} = 1$ and $\alpha$ is restart probability. 
 A RWR starts at $u$ can be seen as a particle starts at $u$ and in each step,
it moves to its neighbours with probability that is proportional to their weight and also returns to $u$ with a probability $\alpha$. 
RWR stops after maximum $m$ steps or when it converges with threshold $\epsilon$.
If RWR stops at iteration $i$, $r = r_{i}$.

We apply RWR as follows:
\begin{itemize}
\item For a target vertex $u$, build a local graph $G_{u} = (V_{u}, E_{u})$ in which $V_{u} = \{u\} \cup N_{u} \cup C_{u}$ and $E_{u}$ is the edges set containing those edges $e_{v, w}, v, w \in V_{u}$
\item Run RWR at $u$ on $G_{u}$
\item When RWR stops, set $\psi_{u, v} = r_{v}$
\end{itemize}
%So $\psi_{u, v}$ can be considered as ability for $u$ to connect to $v$ after a period of activity on the network. 

After calculating scores, the candidates are ranked in descending order by the scores.

\section{Experiments and Discussions}

\subsection{The Used Datasets}
Experiments are performed to evaluate effectiveness of the proposed friend suggestion approach compared to some other approaches: the one currently deployed in ZingMe, Adamic-Adar score, number of mutual friends, and plain Random Walk with Restarts.
The data used in these experiments are obtained from the real data collected by ZingMe, including:
\begin{itemize}
\item The list of new friendships established in ZingMe from August 08, 2013 to October 02, 2013 that consists $3,540,624$ users and $26,040,831$ new friendships.
\item List of friends, schools, groups, and companies of users appeared in new friendships above.
\end{itemize}
We note that many features missed in the collected data, such as the number of mutual IP addresses and the number of mutual interactions. This may make negative effects on suggestion efficiency of the proposed approach. 

Users who have number of new friendships smaller than $5$ are eliminated from dataset in order to keep more active users for evaluation. 
The set of remaining users is called $U$. 
Then, for each user $u \in U$, we remove from friend list of $u$ users who are not in $U$ or make friends with $u$ in the period of data collection. 
Finally, U consists $928300$ users and on average, each user has $90$ friends and $25$ new friendships.
We consider $U$, refined lists of friends, and lists of schools and groups of each $u \in U$ as the snapshot of the network that is used to calculate suggestions for tested users.
%After that, if $u' \in N_{u}$ but $u \notin N_{u'}$, we add $u$ to $N_{u'}$. 
%This happened because of some . So we do this to keep friendships symmetric. 

In order to evaluate performance of approaches to users with different numbers of friends,
we randomly choose three user sets $T20$, $T50$, and $T100$ consisting of about $1000$ users having number of friends from $20$ to $30$, from $50$ to $60$, and at least $100$, respectively.

\subsection{The Experiment Design}
We divide new friendships into cross-validation set consisting ones from August 08, 2013 to August 27, 2013, and test set consisting ones from August 28, 2013 to October 02, 2013.
We estimate parameters or weights of the proposed approach through the cross-validation set. 

We could only collect data for three features: the number of mutual friends, the number of mutual schools and the number of mutual groups. 
Therefore, we set weights of other ones to $0$. 
We change weights of those three features by step $0.1$, keep weight for the number of mutual friends higher than the others and finally get the weights for them equal to $0.5$, $0.3$ and $0.2$, respectively.
Through estimation, we set $w_{d}$ , $w_{i}$ to $0.4$ and $0.6$. 

For RWR, $L = 10000$ helps RWR run in acceptable time from $1$ to $2$ seconds per users on average.
$\mu = 5$ outputs each $u \in T20$ about $100$ suggestions and much higher for users in $T50$ and $T100$.
We set $c = 0.4$, $\epsilon = 10^{-4}$ and $m = 50$. 

%From $U$, we choose randomly 1000 users who have from 20 to 30 friends and 1000 users who have from 50 to 60 friends to evaluate effectiveness of each approach on users with different number of friends. We call these set $T20$ and $T50$, respectively. Suggestions for users in these sets are calculated with graph from $U$. 

%Experiment results are also evaluated over 30000 random users who appear in all four datasets in order to evaluate approaches in graphs with different minimum degrees. Because these users appear in all four dataset, they have at least 100 friends. 

%We choose randomly 100 users in this 30000 users and execute proposed approach with unlimited candidates, ie. $L = \infty$.	

We obtain parameters of the current approach from the real system and present them in Table~\ref{tb:paracurrent}.
The approaches using Adamic-Adar score and number of mutual friends consider all friends of friends of user $u$ as candidates for $u$.
The RWR approach uses parameters as in the proposed approach but it sets strength of all edges to $1$.

\begin{table}[!htb]
    \caption{Values of parameters of current approach}
    \label{tb:paracurrent}
   \begin{subtable}{.5\linewidth}
      \centering
        \caption{Values of $t$}
        \begin{tabular}{ c | c | c | c}
            $t_{1}$ & $t_{2}$ & $t_{3}$ & $t_{4}$ \\ \hline
              1.7     &   1.5      &    1.4    &  1.1
        \end{tabular}
    \end{subtable}%
    \begin{subtable}{.5\linewidth}
      \centering
        \caption{Values of $e$}
        \begin{tabular}{ c || c | c | c | c}
               & 1        & 2       & 3      &    4 \\ \hline \hline
           1  & 0        & 2       & 1.9   &    1.6 \\ \hline
           2  & 2        & 0       & 1.8   &    1.7 \\ \hline
           3  & 1.9     & 1.8    & 0      &    1.4 \\ \hline
           4  & 1.6     & 1.7    & 1.4   &    0 \\ %\hline
        \end{tabular}
    \end{subtable} 
\end{table}

\textbf{The Used Metrics.}
We use \textit{precision} and the Area under the ROC curve (AUC) as the metrics to evaluate the system's performance on friend suggestion.

The precision score indicates how many suggestions for $u$ would actually become friends of $u$.
We measure precision at top $100$ suggestions with thresholds from $1$ to $100$. 
The precision score at threshold $k$ is defined as follows: 
\begin{equation}
\textit{P@k} = \frac{\textit{number of actual friends at top k}}{k}
 \end{equation}
We use the average precision scores at each threshold from all tested users as the finally precision score.

The AUC score indicates the expectation that a true suggestion,i.e., a suggestion actually becomes a friend of $u$, is ranked before a false suggestion, i.e., a suggestion does not become a friend of $u$, and is calculated as follows \cite{ling03}:
\begin{equation}
\textit{AUC} = \frac{S_{0} - n_{0}(n_{0} + 1)/2}{n_{0} n_{1}}
 \end{equation}
 in which $n_{0}$ and $n_{1}$ are the numbers of true and false suggestions, respectively, and $S_{0} = \sum_{i} r_{i}$ where $r_{i}$ is the rank of $i^{th}$ true suggestion in the ranked list.
The final AUC score is calculated by the average of AUC scores of all tested users.
We only measure AUC for the proposed approach and RWR one, and measure whole suggestion lists from these approaches.

\subsection{The Experimental Results and Discussions}
\begin{figure}
        \centering
        \begin{subfigure}[b]{0.32\textwidth}
                \includegraphics[width=\textwidth]{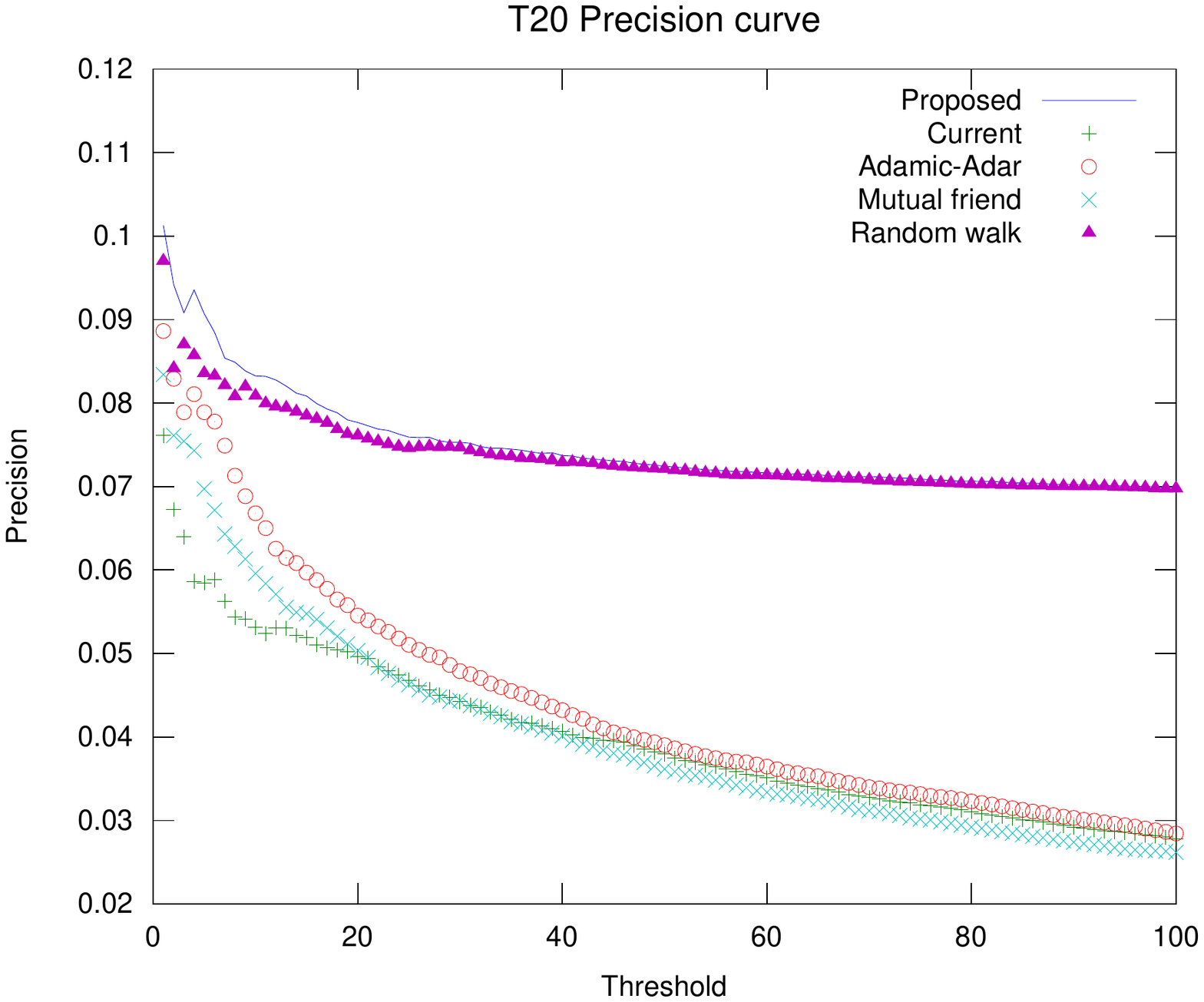}
                \caption{The precision curve for \textit{T20}}
                \label{fig:precision20}
        \end{subfigure}%
        ~%add desired spacing between images, e. g. ~, \quad, \qquad etc.
        \begin{subfigure}[b]{0.32\textwidth}
                \includegraphics[width=\textwidth]{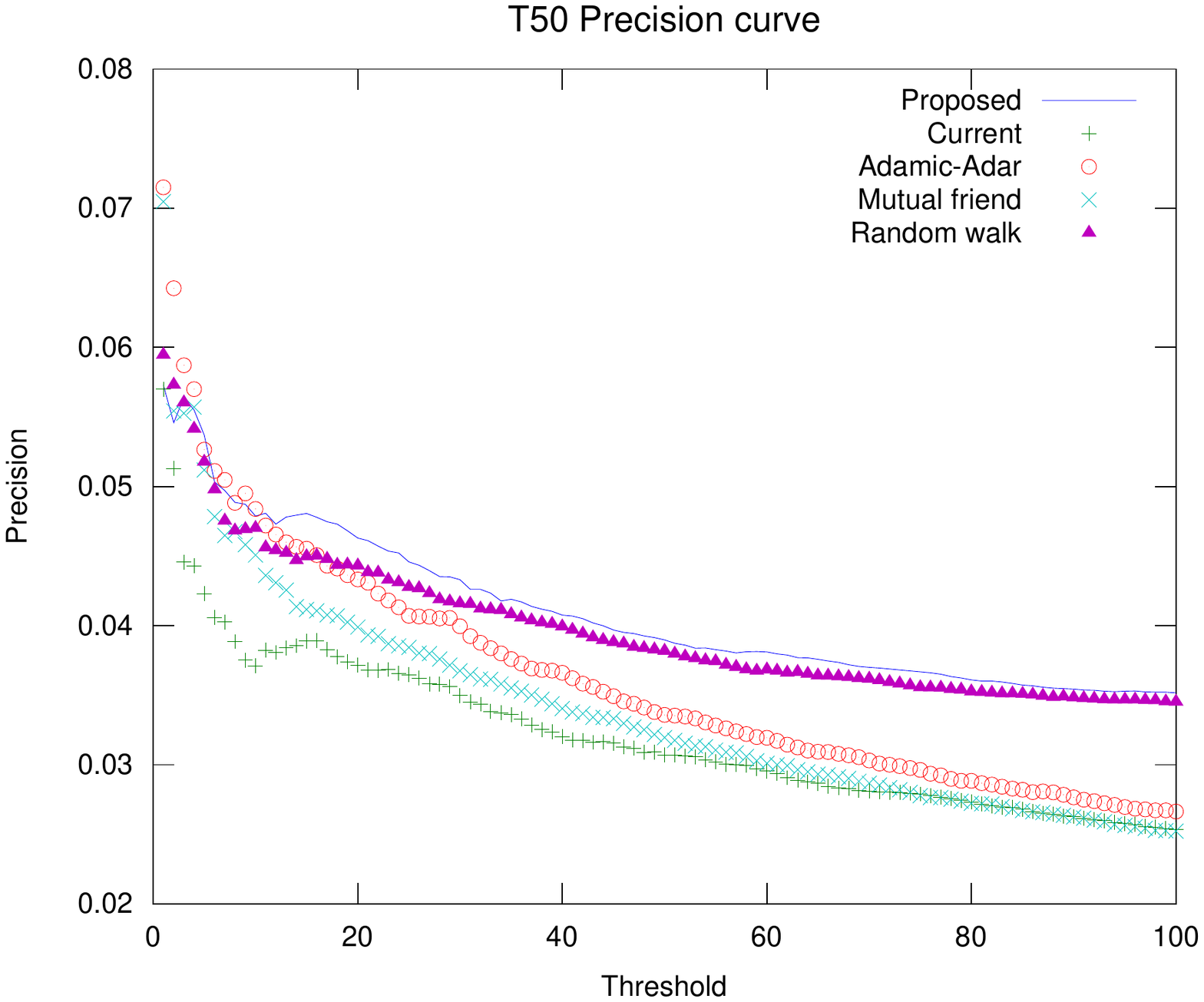}
                \caption{The precision curve for \textit{T50}}
                \label{fig:precision50}
        \end{subfigure}
        ~%add desired spacing between images, e. g. ~, \quad, \qquad etc.
        \begin{subfigure}[b]{0.32\textwidth}
                \includegraphics[width=\textwidth]{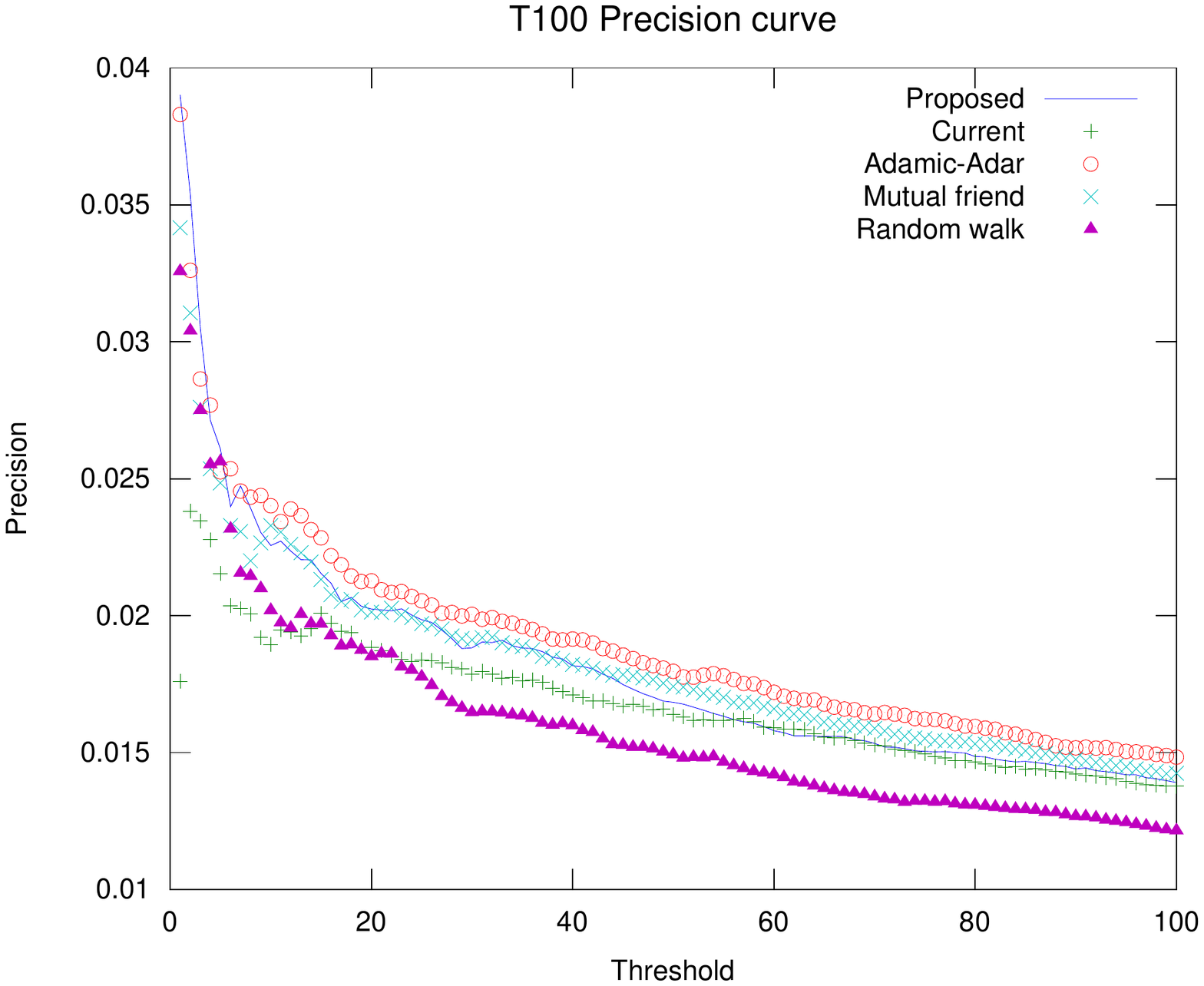}
                \caption{The precision curve for \textit{T100}}
                \label{fig:precision100}
        \end{subfigure}
        \caption{The precision curves of the approaches for \textit{T20}, \textit{T50}, and \textit{T100} user sets with precision threshold from $1$ to $100$}
        \label{fig:precisions}
\end{figure}

Figure~\ref{fig:precisions} shows precision curves of the approaches for different user sets. 
We can see that in all cases, the proposed approach outperforms the current one. 
On average, the improvement is about $98\%$, $28\%$, and $7\%$ in $T20$, $T50$, and $T100$, respectively.

In $T20$ and $T50$, the proposed approach also outperforms Adamic-Adar and number of mutual friends approaches with improvements, on average, about $85\%$  in $T20$ and $20\%$ in $T50$. 
Its performance is slightly lower than those two approaches in $T100$.

Table~\ref{tb:auc} shows AUC scores of the proposed and plain RWR approaches. The proposed approach performs better than plain RWR one in both precision and AUC scores.

%Results of testing same users through different datasets presented in Table \ref{tb:resultsameusers} also shows the improvement of proposed approach over other approaches on datasets with different properties.

\begin{table}[t]
\begin{center}
\caption{AUC scores of the proposed and the RWR approaches in all user sets}
\label{tb:auc}
	\begin{tabular}{c | c | c}
	%\hline
	User set        & Proposed          & RWR \\ \hline \hline
	\textit{T20}    & \textbf{0.599}    & 0.589    \\ \hline
	\textit{T50}    & \textbf{0.620}    & 0.602    \\ \hline
	\textit{T100}  & \textbf{0.626}    & 0.582    \\ %\hline
	\end{tabular}
\end{center}
\end{table}

Our proposed approach outperforms other ones in almost all of cases. However, when the number of friends of target user $u$ increases, the improvement tends to decrease. 

This can be explained by missing features in datasets. The proposed approach uses more features than others but those features are not existed in datasets. 
They are the number of mutual IP addresses and the number of mutual interactions. 
Especially, when the number of friends of $u$ increases, the number of mutual interactions between $u$ and $v$ may represent $\textit{aff}_{u, v}$ clearer than some other features under the hypothesis that the more number of mutual interactions two users have, the more affined they are.

Moreover, the increase in the number of friends often leads to drastic increase in the number of friends of friends of $u$. Then, with limitation $L$ for the number of candidates, the proposed approach may not cover enough potential candidates. Meanwhile, all other approaches, except plain RWR one, calculate suggestion score for all friends of friends of $u$.

By evaluating the approaches through different sets of users with different number of friends,
we can see that the proposed approach performs better when users have smaller friends. 
Besides the reason from limitation $L$,
another reason may be that increasing number of friends of users may reduce differences among candidates,i.e., differences among calculated affinities, and then cause the approach more difficulty to differentiate them. 
 For example, with number of mutual friends feature only, if two candidates $v_{1}$ and $v_{2}$ have $5$ and $6$ mutual friends with $u$, respectively, difference between them is $|0.5 * \textit{log}(6) - 0.5 * \textit{log}(7)| = 0.077$. 
 When number of their friends increase, number of mutual friends with $u$ may increase. If they have $99$ and $100$ mutual friends with $u$, difference is $|0.5 * \textit{log}(100) - 0.5* \textit{log}(101)| = 0.005$. 
 This experimental result may indicate that the proposed approach should be used at the beginning use period of the networks or for those users with small number of friends.

\section{Related Works}
The link prediction problem for social networks was formally defined by Liben-Nowell and Kleinberg in \cite{liben03}. They proposed and compared some approaches for this problem, such as shortest path, mutual neighbors, Jaccard, Adarmic-Adar and so on. They evaluated on social networks of co-authorships via scientific papers. The effectiveness was evaluated over a random approach. These approaches only leverage network structure information. 

There are many works on link prediction for social networks using only network structure information for prediction new connections. Huang \cite{huang06} used information from clusters in the network. %Kunegis et al. [2] used graph kernels for bipartite networks. There was no optimal approach for all of their datasets. 
Rattigan and Jensen \cite{rattigan05} focused on a smaller problem: anomalous link detection. They showed effectiveness of applying link prediction models for anomalous link detection problem. They also presented a fundamental problem causing low raw results of link prediction model. It is a massive class skew. %Symeonidis et al. [4] presented many different features but they only were derived from network structures. 
Menon and Elkan \cite{menon11} used matrix factorization and showed that their approach can combine more types of information. %Matrix factorization was also used by Kunegis and Lommatzsch [8]. 
Sun et al. \cite{sun12} exploited network structure to solve link prediction problem in heterogeneous networks. Not only appearance of connections, they also predicted time of the appearance. Some works \cite{mislove08}, \cite{omad05} %, [10] 
solved this problem on growing networks instead of static networks.

One of reasons making most of works focus on network structure is difficulty in obtaining other kinds of information of users, such as attributes of users or their interactions. In real systems, these kinds of information can be obtained but user privacy should be considered. Machanavajjhala et al. \cite{macha11} presented an approach for a compromise between suggestion accuracy and user privacy. 

In addition to network structure, some works used network structure for the link prediction problem.
Yin et al. \cite{yin11} presented some factors that may effect on establishing connections and make it easier to connection to this user than that one. They showed that random walk was suitable for this problem and satisfied those factors. %Zheleva et al. [13] used information from network structure, node attributes and group of nodes. 
%Hasan et al. \cite{hasan06} applied some supervised approaches such decision tree, SVM and so on. 
%Wang et al. \cite{wang07} also used supervised approaches by building a probabilistic graphic model and showed that that could help them obtain features, which were difficult to be obtained from network structure and attributes of nodes. 
De et al. \cite{de12} hypothesized that connection occurrence between two nodes depended only on local properties. 
%Schifanella et al. [16] used tags users provided for content management and sharing systems. 
Scellato et al. \cite{scell11} and Wang et al. \cite{wang11} used location and mobility of nodes. For social networks that have these types of information, leveraging them may be a good direction. 

Backstrom and Leskovec \cite{back11} exploited network structures, attributes of nodes and edges for this problem. They presented supervised random walk that guided random walk so that the random walker was more likely to visit a node that would connect to target node. But they did not use direct information between two nodes for prediction. 
Tylenda et al. \cite{tylen09} also used these types of information for growing networks. In addition, in their approach, adjacent vertices did not need to be connected. This may help choosing adjacent vertices more flexible. 

RWR is a well-known technique for proximity measure on graph \cite{tong06}, \cite{sarkar09}, %\cite{tong06b}, 
\cite{tong07}.
% It is also used to rank nodes on graph \cite{agar07}, \cite{agar06}, \cite{dili05}, \cite{minkov07}.

\section{Conclusions and Future Works}
With the goal of helping users of online social networks build and extend their relationships by suggesting to those users new friendships, we presented some formal definitions of the link prediction problem, the current approach of ZingMe and proposed a new approach for this social network. 
The proposed approach exploits knowledge sources from network structure, attributes of nodes and edges to archive direct information between two users and indirect information from other users in the network. These types of information are used to assess ability to establish friendships among users. Random walk with restart is the method used to calculate indirect type of information. 

Some experiments were conducted to evaluate effectiveness of the proposed approach on a big dataset collected from ZingMe. The proposed, current and some other well-known approaches were evaluated on this dataset. The results from proposed approach have exceeded other ones. In some cases, a poor improvement of current approach over other ones can be explained by the missing of features in the dataset. We also suggest proposed approach for users who have small number of friends and who the proposed approach can output much better suggestion lists compared to other ones. 

There are still many knowledge sources in ZingMe ignored, which may be useful for friend suggestion problem. For example, chat rooms a user has entered, games a user has played and so on. Collecting them and finding more sources should be considered. 

Interest of users to suggestions may vary in different contexts. For example, when a user $u$ visits profile page of a user $v$, $u$ may incline to make friends with friends of $v$ at that time. So we can consider a context-aware recommendation approach.

There are many fixed parameters in the proposed approach, for example, the weights for the features in Equation~\ref{eq:aff}, or $w_{d}$ and $w_{i}$ in Equation~\ref{eq:score}. A model to learn them may improve performance. 

User feedbacks to presented suggestions can be a good knowledge source to remove bad suggestions and to avoid repeating suggestions to users. For instance, if a user $v$ is suggested to $u$ many times but $u$ has not made friend with $v$, perhaps $v$ is not a good suggestion for $u$ although $\textit{score}_{u, v}$ might be high.

%For friend suggestion problem, we only predict new friendships likely to exist in the future. In real social networks, suggestions should also attract users. It sometimes needs a trade-off between novelty and accuracy. 

\end{document}